%% file: main.tex
\documentclass[sigconf]{acmart}

\usepackage{subcaption}

\usepackage{multirow}

\usepackage{listings}
\lstdefinestyle{listingstyle}{
    basicstyle=\ttfamily\footnotesize,        
    breaklines=true,       
    frame=lines,
    numbers=none
}
\usepackage{xspace}
\newcommand{\approach}{{$FeedAIde$}\xspace}

\usepackage{tabularx}

\usepackage{enumitem}

\AtBeginDocument{%
  }

\copyrightyear{2026}
\acmYear{2026}
\acmConference[MOBILESoft '26]{IEEE/ACM 9th International Conference on Mobile Software Engineering and Systems}{April 12, 2026}{Rio de Janeiro, Brazil}
\acmBooktitle{IEEE/ACM 9th International Conference on Mobile Software Engineering and Systems (MOBILESoft '26), April 12, 2026, Rio de Janeiro, Brazil}
\acmPrice{}
\acmDOI{10.1145/3795077.3795120}
\acmISBN{979-8-4007-2486-2/2026/04}

\begin{document}

\title{$FeedAIde$: Guiding App Users to Submit Rich Feedback Reports by Asking Context-Aware Follow-Up Questions}


\author{Ali Ebrahimi Pourasad}
\email{ali.ebrahimi.pourasad@uni-hamburg.de}
\orcid{0009-0008-7206-726X}
\affiliation{
  \institution{University of Hamburg}
  \city{Hamburg}
  \country{Germany}
}

\author{Meyssam Saghiri}
\email{meyssam.saghiri@studium.uni-hamburg.de}
\orcid{}
\affiliation{%
  \institution{University of Hamburg}
  \city{Hamburg}
  \country{Germany}
}

\author{Walid Maalej}
\email{walid.maalej@uni-hamburg.de}
\orcid{0000-0002-6899-4393}
\affiliation{
  \institution{University of Hamburg}
  \city{Hamburg}
  \country{Germany}
}
\renewcommand{\shortauthors}{Ebrahimi Pourasad et al.}

\input{chapters/00_abstract}




\maketitle

\input{chapters/01_introduction}
\input{chapters/02_approach}
\input{chapters/03_approach_implementation}
\input{chapters/04_evaluationDesign}
\input{chapters/05_evaluationResults}
\input{chapters/06_threats}
\input{chapters/07_relatedWork}
\input{chapters/08_discussionAndConclusion}

\balance
\bibliographystyle{ACM-Reference-Format}
\bibliography{references}

\newpage
\input{chapters/09_appendix}

\end{document}

%% file: chapters/00_abstract.tex
\begin{abstract}
User feedback is essential for the success of mobile apps, yet what users report and what developers need often diverge.
Research shows that users often submit vague feedback and omit essential contextual details. 
This leads to incomplete reports and time-consuming clarification discussions. 
To overcome this challenge, we propose \approach, a context-aware, interactive feedback approach that supports users during the reporting process by leveraging the reasoning capabilities of Multimodal Large Language Models. 
\approach captures contextual information, such as the screenshot where the issue emerges, and uses it for adaptive follow-up questions to collaboratively refine with the user a rich feedback report that contains information relevant to developers.
We implemented an iOS framework of \approach and evaluated it on a gym’s app with its users.
Compared to the app’s simple feedback form, participants rated \approach as easier and more helpful for reporting feedback. 
An assessment by two industry experts of the resulting 54 reports showed that \approach improved the quality of both bug reports and feature requests, particularly in terms of completeness.
The findings of our study demonstrate the potential of context-aware, GenAI-powered feedback reporting to enhance the experience for users and increase the information value for developers.
\end{abstract}

%% file: chapters/01_introduction.tex
\section{Introduction} \label{sec:introduction}

User satisfaction is a critical factor driving the success of mobile apps in an increasingly competitive market \cite{Martens:ReleaseEarlyRelease:2019,Lee:DeterminantsMobileApps:2014,pagano_user_2013}.
Consequently, user feedback is essential for developers to understand user needs and improve their apps, for example, by fixing reported bugs or adding requested features \cite{maalej2025automated,Martens:RE:19}.
App stores enable users to easily submit feedback as a text review and a 1-5 star rating.
While user feedback has become very popular in app stores with some apps receiving thousands of reports per day \cite{pagano_user_2013}, prior work shows that much of the feedback submitted is rather uninformative, vague, or purely negative \cite{Martens:RE:19,pagano_user_2013}.
Non-technical users often provide unstructured reports that lack essential contextual details \cite{Martens:RE:19}.
This creates a gap between what users report and what developers need \cite{Hassan:EMSE:2018}, potentially leading to user frustration if their concerns appear not to be taken seriously \cite{Lee:DeterminantsMobileApps:2014}.
Research shows that about 45\% of developer responses to app store reviews ask for more details.
Many users never reply to follow-up questions, and those who do often take several hours or days \cite{Hassan:EMSE:2018,breu2010information}.
Prior work has proposed enhancing App Stores to increase the overall quality of apps by better supporting developers \cite{gomez2017app}.
These studies highlight that user feedback from app stores is valuable but often incomplete, leading to time-consuming clarification rounds.

Advances in Generative AI (GenAI), particularly Multimodal Large Language Models (MLLMs), have revealed powerful dialogue and context reasoning capabilities that can assist the gathering of higher quality user feedback for app development \cite{Pourasad:DoesGenAIMake:2025, wei2024getting, wei2025guing}.
Recent research has explored how GenAI can help developers analyze and enhance user feedback \cite{acharya2025can, du2024llm, maalej2025automated}.
From the user perspective, prior approaches have also attempted to support feedback reporting \cite{seyff2014appecho, song2022toward}, but none have utilized context data (such as device information, screenshot and interaction logs) together with GenAI capabilities to instantly guide users and simplify their reporting experience while creating more valuable, rich feedback for developers.

To address this gap, we introduce \approach, a novel feedback approach that enables app users to easily report feedback by answering \textbf{adaptive follow-up questions} in order to gather information needed by developers.
When a user wants to submit feedback, \approach automatically captures contextual information, such as a screenshot, and uses this context to propose several feedback options for the user to choose from.
It then refines the chosen option together with the user through follow-up questions to create a detailed feedback report.
Building on the generalization capabilities of MLLMs, \approach is not restricted to a single type of feedback, such as bug reports or feature requests, nor to a specific app domain. 
Instead, it can handle diverse feedback scenarios, with follow-up questions dynamically created based on the captured context. 
Section~\ref{sec:approach} presents the approach and an iOS implementation, which constitutes our primary contribution.  

To evaluate \approach, we conducted a study on a production gym app.
We describe the study design in Section~\ref{sec:eval-design}.
To assess whether \approach provides an easy and helpful experience to users when submitting feedback, we performed user testing with actual app users.
The participants simulated reporting four previously submitted feedback cases, once with \approach and once with the app’s simple text-field form.
Afterward, we surveyed participants to understand their perception on \approach.  
The results showed that users preferred \approach over the app’s standard feedback form, describing it as easier and more helpful to use, while suggesting improvements such as reduced loading times.
To evaluate the value of FeedAIde for developers, two industry experts assessed 54 reports collected during the user testing.
Their assessment confirms that \approach produced higher-quality reports for both bug reports and feature requests, while highlighting potential areas for improvement, such as adjusting follow-up questions to uncover the root cause of an issue rather than refining a solution.
Section~\ref{sec:eval-results} reports the evaluation results, which together with the data available in our replication package \cite{replicationPackage}, constitute our second contribution.
The remainder of the paper discusses threats to validity in Section~\ref{sec:threats}, reviews related work in Section~\ref{section:related_work}, and concludes with a summary and future work on GenAI-assisted user feedback flows in Section~\ref{sec:conclusion}.

%% file: chapters/02_approach.tex
\section{Approach} \label{sec:approach}

We propose \approach, a context-aware feedback approach that leverages MLLMs to guide users in creating rich feedback reports.
The framework depicted in Figure~\ref{fig:flow} collects contextual information, such as recent user interactions and a screenshot, to understand the user’s context.
When the user triggers the feedback system on their mobile device, for example through a shake gesture, \approach uses this context to prompt an MLLM to generate concise, context-dependent, textual feedback predictions.
The user chooses one or manually writes their feedback and the model follows up with adaptive questions to gather missing details.
For example, when a user cannot log in after several failed attempts, the system detects the repeated actions and proposes ``Report a login problem''.
The follow-up questions might ask whether the user recently changed their password and whether the log-in works on another device, ensuring the developer receives a deeper understanding of the reported issue.
Finally, the model creates a rich and understandable feedback report containing the gathered data and a summary.

In the following, we present the workflow of \approach and an example based on our iOS framework implementation.

\subsection{Workflow}

\begin{figure}
    \centering
    \includegraphics[width=1.0\linewidth]{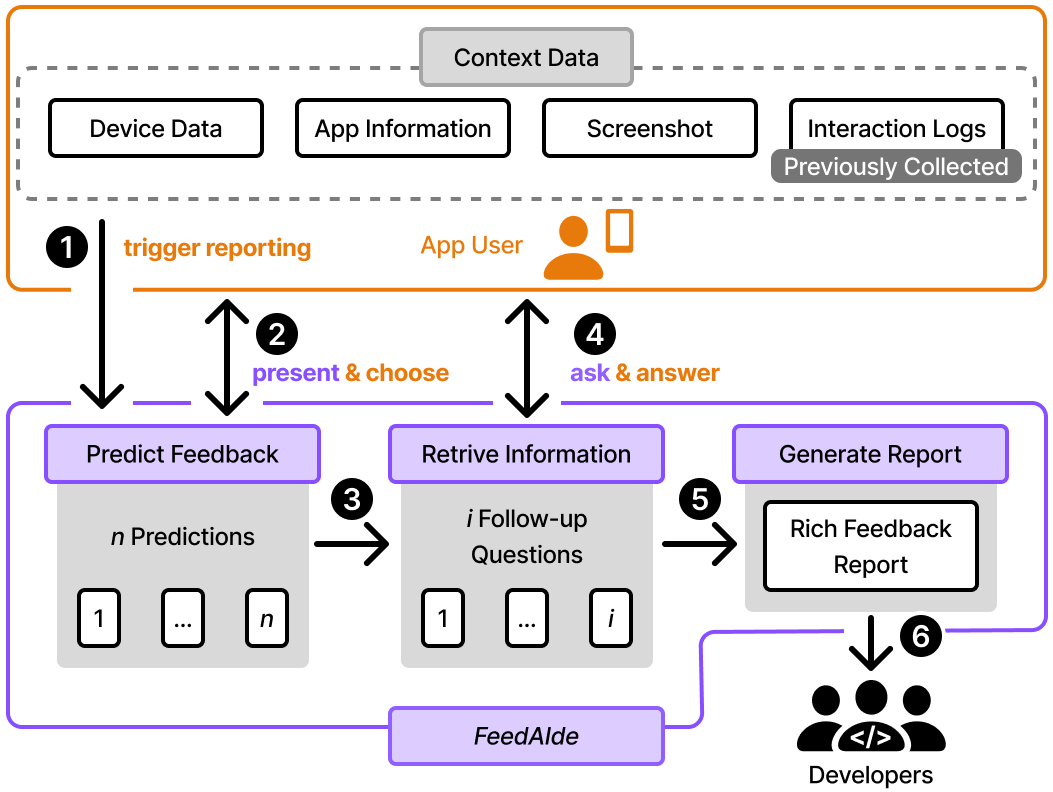}
    \caption{
Overview of \approach.  
When a user triggers the feedback process, \approach collects contextual data, such as a screenshot (1).
\approach then predicts possible feedback, such as a bug report, from which the user can choose or enter their own (2).  
Next, \approach generates $i$ context-aware follow-up questions (3), asks the user for answers to gather additional details (4), combines all information into a rich feedback report (5), and sends it to the developers (6).
}
    \label{fig:flow}
\end{figure}

Figure~\ref{fig:flow} illustrates the overall workflow of the \approach concept.  
When a user triggers the feedback reporting process, \approach gathers contextual data, which consist of a screenshot, device details, app information, and previously collected interaction logs (1).
\approach then generates $n$ feedback predictions representing what the user might want to report, allowing the user to choose one or write their own feedback (2).
These predictions can cover different types of feedback, such as bug reports or feature requests, like ``Add a dark mode option''.
Next (3), \approach asks a series of $i$ context-aware follow-up questions to gather additional details and enrich the feedback report (4).  
It then combines all gathered data and user responses into a rich feedback report that includes a summary, a history of all questions and user answers, as well as the collected contextual data (5).
Finally, the completed feedback report is sent to the developers in a JSON format (6).

\begin{figure*}
\centering
\includegraphics[width=1.0\textwidth]{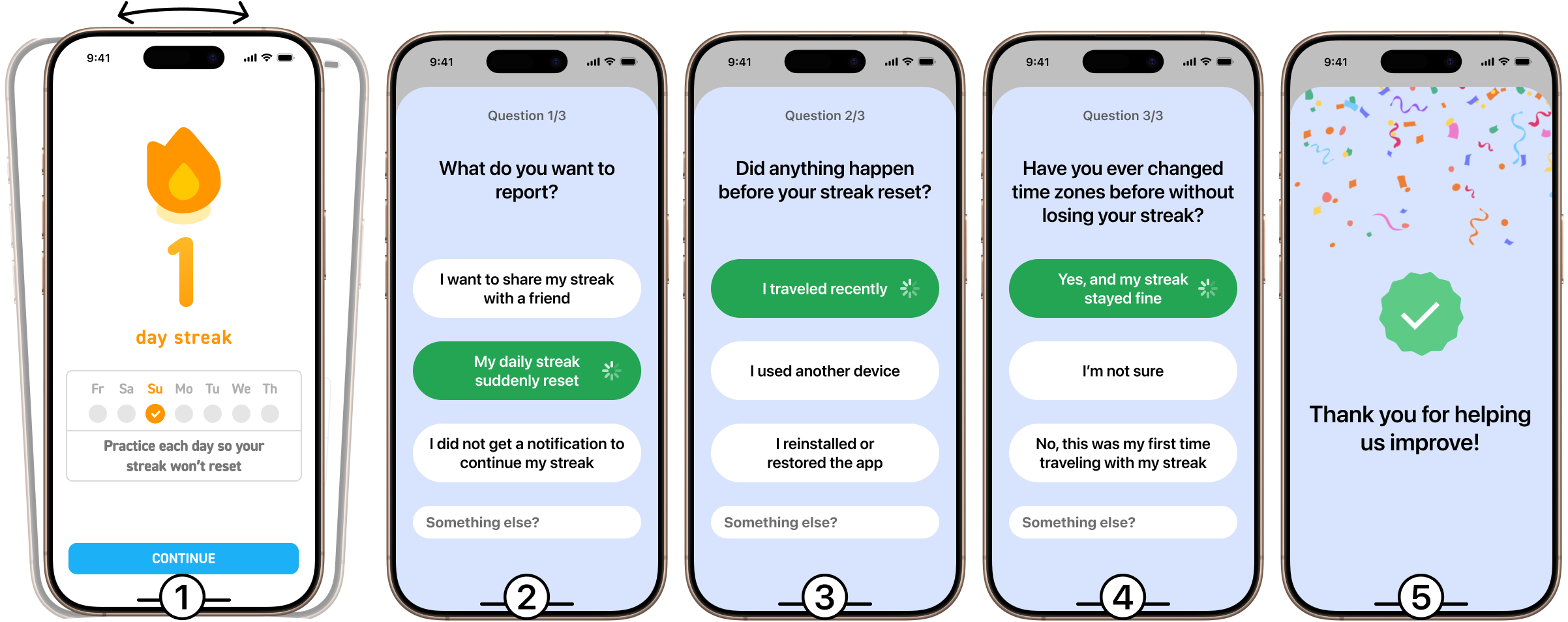}
\caption{
\approach iOS framework guiding users through a context-aware feedback flow. 
In this example, a user reports that their daily streak in a language learning app has unexpectedly reset to one (Screen~1).
Once triggered, e.g. via shake-to-report, \approach detects the context and proposes possible feedback (Screen~2).
After the user selects ``My daily streak suddenly reset'', the system asks adaptive follow-up questions to collect valuable information for the developers (Screens~3–4).
It then creates a rich feedback report and confirms its submission (Screen~5).
}
\label{fig:ios-approach}
\end{figure*}

An example from our iOS framework implementation of \approach is shown in Figure~\ref{fig:ios-approach}.
It shows a language learning app, where a daily learning streak unexpectedly resets to one (Screen~1).
When \approach is triggered by the user, for instance by shaking the device or pressing a ``Report Feedback'' button, it processes the previously described context data using an MLLM.
Based on this information, the model tries to predict what the user might want to report and presents three feedback options (Screen~2).
Here, the user selects ``My daily streak suddenly reset'', which initiates the generation of adaptive questions (Screens~3-4).
Through two follow-up questions, \approach gathers additional details that could help with explaining the underlying issue while keeping the interaction minimal for the user.
In this example, \approach detects from the interaction logs that the activity timestamps differ across time zones and uses this insight to ask targeted follow-up questions to clarify whether the streak reset might be related to recent travel.
After gathering the user’s responses, the system generates a rich feedback report and sends it to the developers while displaying a successful submission view to the user (Screen~5).
The final report contains a summary, contextual information including device information, as well as the answers to the adaptive questions.

%% file: chapters/03_approach_implementation.tex

In the following, we present the implementation details, including the architecture, model, and prompt design.

\subsection{iOS Implementation}
\label{sec:implementation}

We implemented \approach as a Swift Package for iOS applications, which is a reusable bundle of Swift code that can be easily integrated into other projects~\cite{apple_swift_packages_doc}.
The UI is built using Apple's declarative UI framework, SwiftUI~\cite{swiftui}, and can be displayed in any app by calling a single view modifier.
After adding this top-level modifier, the feedback flow becomes fully functional and automatically collects contextual information such as a screenshot and device data.
However, developers still need to integrate specific logging events in their app so that the framework can utilize them.
A \textit{shake to report} gesture is included to trigger the feedback flow, though developers may alternatively implement a standard feedback button to start the process.

To integrate an MLLM while maintaining modularity, low coupling, and easy model switching, we used the AIProxy~\cite{aiproxy} Swift library.
AIProxy serves as a proxy that forwards requests to the respective APIs of major LLM providers such as OpenAI.
This design simplifies implementation and ensures a secure connection by keeping API keys outside the app binary.
Additionally, AIProxy prevents misuse through Apple's Device Check API~\cite{apple_devicecheck}, which verifies that the framework is accessed only from legitimate installations on physical Apple devices, reducing the risk of abuse from simulated or unauthorized environments.

\subsection{Prompt Engineering}  
We first conducted pilot tests with several prompt configurations and then refined them based on the findings, as detailed in the replication package \cite{replicationPackage}.
In short, to mitigate the risk of \textit{oversharing data} with the MLLM, only the most relevant contextual information should be provided.  
Oversharing is undesirable because it increases token consumption, may slow down response times, and create a ``needle in a haystack'' scenario.
That is, excessive context can bury the relevant details and mislead the model \cite{liu2023lost, machlab2024llm}.
Also, only essential contextual data should be transmitted, helping to safeguard user privacy and reduce unnecessary data exposure.
Nevertheless, all essential parameters must be included to enable the model to ask appropriate follow-up questions \cite{shin2025can}. 
These parameters were selected based on testing and insights from prior research, such as work outlining the characteristics of a ``good bug report'' by Bettenburg et al.~\cite{bettenburg2008makes}.

The final prompt includes three sections:  
(1) Setting, Instructions, and Constraints, 
(2) App Description, and  
(3) Context Information.  


\begin{lstlisting}[float, numbers=none, caption={Settings Prompt}, label={lst:promptSetting}]
# Setting
You are a specialized LLM that functions solely as an intelligent, low-friction user feedback system in an iOS app. You will be used for all types of feedback including bugs, suggestions, and content complaints. The user is unaware that they are communicating with an LLM. Always adhere to the stated protocol!
The goal is that developers receive helpful, context-specific feedback based on actual user behaviour to avoid frustrating users. The questions asked should both address the user's concerns and be useful for the developer. This aims to minimize asynchronous communication.

You will receive a screenshot of the current screen and the initial interaction context of the feedback flow in the following JSON format:
```json
{
  "events": ["app_launched", "login_button_pressed"],
  "eventTimestamps": ["2025-05-09T16:23:00+02:00", "2025-05-09T16:24:00+02:00"],
  "feedbackInitiationTimestamp": "2025-05-09T16:23:00+02:00",
  "appVersion": "1.3.2",
  "deviceInfo": {
    "model": "iPhone13,4",
    "osVersion": "iOS 17.5",
    "language": "de"
  }
}
```
\end{lstlisting}

\subsubsection{Setting, Instructions, and Constraints} 
This section (1) is identical across all apps and defines the behavior of the model.
Listing~\ref{lst:promptSetting} shows the ``Setting'' part of the prompt, which defines the LLM’s role and overall objective.  
According to Shin et al.~\cite{shin2025can}, role-prompting is an effective approach to define the model’s behavior.
With role-prompting, the LLM is instructed to take the role of a system whose sole purpose is to function as a feedback system.
First, the role of the LLM is described, emphasizing that it must always adhere to the defined protocols.
Furthermore, the user should not perceive the interaction as a dialogue with an AI system, as such experiences might be linked to negative user perceptions \cite{khan2024consumers}.
Next, the goal of the system is explained, which is to prevent user frustration \cite{bettenburg2008makes} and to reduce asynchronous communication \cite{Martens:RE:19, Hassan:EMSE:2018}.  
Finally, the input that the model will receive is defined.

\begin{lstlisting}[float, numbers=none, caption={Instruction Prompt}, label={lst:promptInstructions}]
# Instructions
Write in the language that is specified in the `deviceInfo.language` field.
Your task is to firstly propose predictions of what the user might want to report based on the initial context. Then you will ask follow-up questions.
The sequence of events is as follows:

1. You will receive the initial context. Analyze the context and generate up to three concrete feedback predictions (`feedbackPredictions`) that the user can select from.
2. After you receive a response from the user, which will either be one of the `feedbackPredictions` or something else, analyze the user's response and send an appropriate follow-up question with up to three concise answering options.
3. After you receive a response from the user, which will either be one of the options or something else, analyze the user's response. Send another appropriate follow-up question with up to three concise answering options.
4. Final Step: After you receive a response from the user, which will either be one of the options or something else, leave `followUpQuestion = null`. Finalize the flow by filling all other fields.

To sum it up:
- 3 predictions
- one question
- one question
- summary
\end{lstlisting}

After the setting is set, the LLM receives its instructions as shown in Listing~\ref{lst:promptInstructions}.  
These describe how the LLM should process input, generate predictions, and guide the flow until the final issue report is produced.

\begin{lstlisting}[float, numbers=none, caption={Constraints Prompt}, label={lst:promptConstraints}]
# Constraints
- The `feedbackPredictions` and answer options should never contain a generic "Something else".
- Make sure all questions are meaningful and relevant. However, they are for non-technical users, so keep them simple and concise.
- The number of follow-up questions you ask is always 2. They have to be concise.
- Never skip ahead to a later step, even if you believe you already know the final answer.
\end{lstlisting}

Next, the model receives a list of constraints, as shown in Listing~\ref{lst:promptConstraints}.  
These constraints were identified during pilot testing, such as answer options containing a generic ``something else'' or questions that were overly technical or too long.  


\subsubsection{App Description}
The App Description section (2) allows developers to describe to the framework their app in detail.  
An example of a shortened description of a language learning app is shown in Listing~\ref{lst:app_description}.
It outlines the app’s overall purpose and key screens.  
The rationale is that a deeper understanding of the app’s structure and functionality enables the MLLM to recognize relevant relationships and produce more accurate feedback predictions, follow-up questions, and issue summaries.

\begin{lstlisting}[float, numbers=none, caption={Example App Description}, label={lst:app_description}]
# App Description
LingoLearn is a mobile app that helps users learn new languages through short, interactive lessons. It focuses on daily practice with vocabulary, grammar, and pronunciation exercises, using streaks and rewards to keep users motivated.

# Description of individual screens
## Home
Shows the current streak, daily goal, and quick access to the next lesson.

[Description of the other screens]
\end{lstlisting}

\subsubsection{Context Information}  
The Context Information (3) is the only dynamic part of the prompt, as it provides details about the user’s current situation within the app.  
A short example can be seen in the JSON block at the end of Listing~\ref{lst:promptSetting}.  
Additionally, a screenshot of the screen from which the report was triggered is added.
Based on this combined contextual information, the MLLM can interpret the user’s context and begin formulating feedback predictions. \\ \\
\begin{figure}
    \centering
    \includegraphics[width=0.9\linewidth]{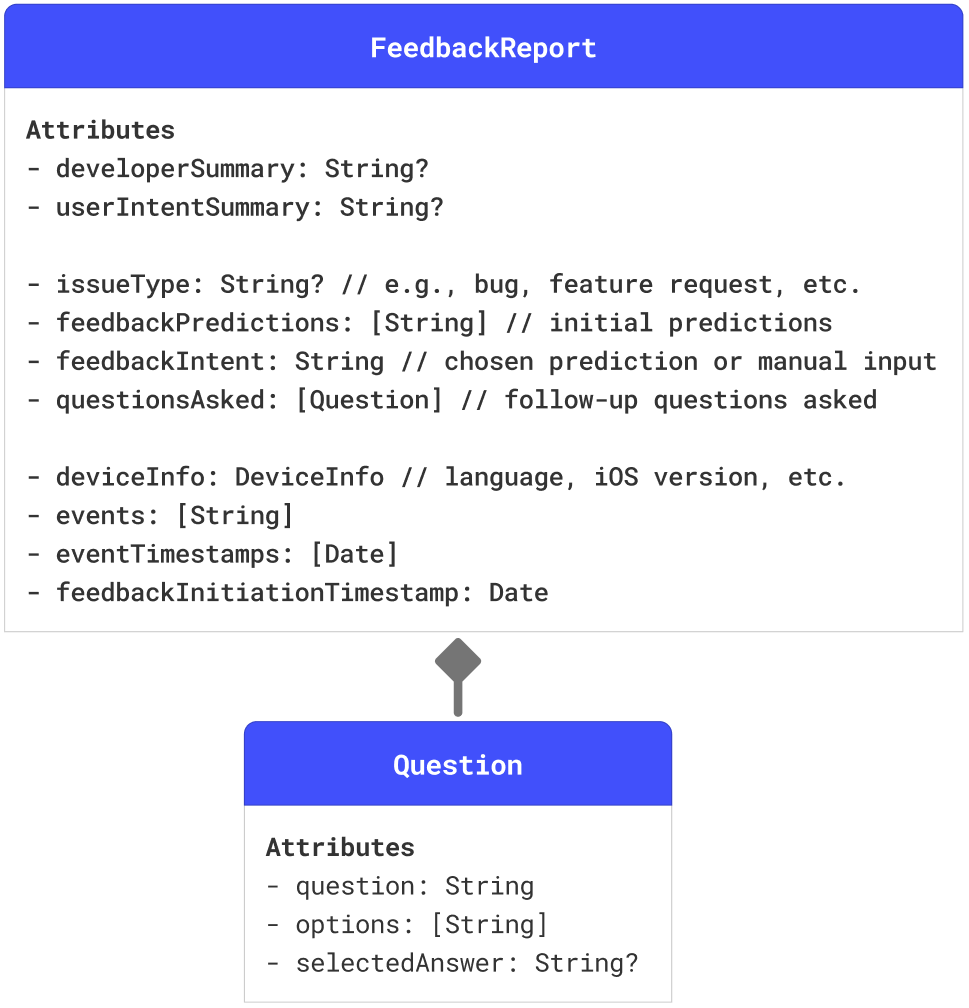}
    \caption{
    UML class diagram of the feedback report generated with \approach.
    }
    \label{fig:uml}
\end{figure}
The model is set up with a temperature of 1 and provided a schema for its structured outputs \cite{openai_structured_outputs_2025}, which constrains the format of its responses. 
This minimizes the need for type-safety checks.
After the model setup, the feedback process unfolds as an interactive sequence between the user and the MLLM.
The system first analyses the contextual information to generate feedback predictions.
Once the user selects or writes one, the model continues with two targeted follow-up questions to clarify and enrich the report.
In the final iteration, the MLLM summarizes the collected input into a structured object, as shown in Figure~\ref{fig:uml}.  
This object contains all collected information, including user interactions, contextual data, and two summaries: \texttt{userIntentSummary} and \texttt{developerSummary}.
The \texttt{userIntentSummary} captures what the user wanted to communicate (e.g., ``My daily streak reset unexpectedly after traveling to another time zone''), while the \texttt{developerSummary} translates this into a concise, developer-oriented form (e.g., ``Potential issue with streak persistence logic related to time zone handling after travel; not observed in earlier versions'').
The object is then converted to JSON and sent to the developers, who can visualize it in their preferred format.

%% file: chapters/04_evaluationDesign.tex
\section{Evaluation Design} \label{sec:eval-design}

To explore the potential and limitations of \approach, we conducted a study guided by the following research questions:

\begin{description}
\item[\hypertarget{rq1}{RQ1}] How do users perceive submitting feedback with \approach compared to a traditional feedback form?
\item[\hypertarget{rq2}{RQ2}] How does the quality of feedback generated with \approach differ from that written through a traditional feedback form?
\end{description}

To answer \hyperlink{rq1}{RQ1}, we used a gym's app to conduct a within-subject user test.
In this testing, participants reported feedback using both the app’s traditional text-based method and \approach.
Based on the 54 collected feedback reports, we conducted an expert assessment to compare their quality and answer \hyperlink{rq2}{RQ2}.
We present the details in the following.

\subsubsection*{Study App}

\begin{figure} 
\centering 
\includegraphics[width=0.9\linewidth]{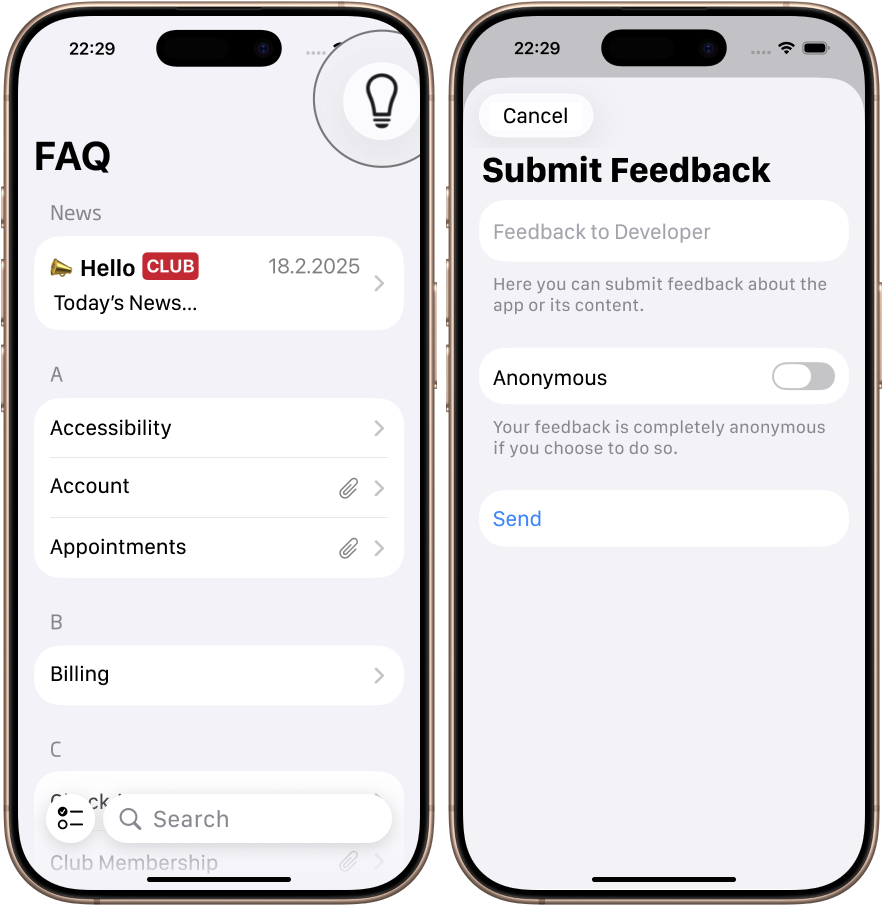} 
\caption{Existing feedback mechanism in the \textit{PPEmployee} app, where users can submit feedback via the FAQ section by tapping the light bulb icon.}
\label{fig:app-old-feedback} 
\end{figure}

\textit{PPEmployee} (pseudonym) is an internal iOS app developed by one of the authors for a local gym.
The app is designed to support employees with their daily operational tasks, such as completing checklists and accessing relevant FAQs.  
It is developed in SwiftUI and distributed internally and is therefore not listed on the App Store.
Its current feedback mechanism, shown in Figure~\ref{fig:app-old-feedback}, allows users to navigate to the FAQ section, tap a light bulb icon, and submit feedback through a simple text field.  
Despite being in use for over a year and having around 20 active users, only two issue reports had been submitted during that period.

\subsection{User Testing (RQ1)}

\begin{table}[b]
    \centering
    \caption{Participants in the user testing.}
    \begin{tabularx}{\linewidth}{|c|c|c|X|}
        \hline
        Participant & Age & Gender & Occupation \\ [0.5ex]
        \hline
        P1 & 21 & M & Fitness trainer \\
        P2 & 51 & M & Club General Manager \\
        P3 & 21 & M & Fitness Trainer \\
        P4 & 25 & F & Fitness Trainer \& Service \\
        P5 & 25 & M & Personal \& Fitness Trainer \\
        P6 & 27 & M & Team Leader \\
        P7 & 19 & M & Fitness Trainer \\
        \hline
    \end{tabularx}
    \label{table:userTestingParticipants}
\end{table}

To assess user perception (\hyperlink{rq1}{RQ1}), we conducted a controlled within-subject user testing using real feedback cases.
We selected four reports, consisting of two bug reports and two feature requests, that had originally been reported by users either verbally or through the app’s feedback feature.
Participants were asked to recreate the same four reports using both feedback mechanisms: the traditional one and \approach.
We recruited seven actual users of \textit{PPEmployee}, whose demographics are shown in Table~\ref{table:userTestingParticipants}.
We mitigated potential order effects by randomly dividing participants into two groups and applying a counterbalancing technique \cite{lazar2017research}.
Each session lasted approximately 20–30 minutes.
The user tests were conducted on-site at the gym’s employee area.
All sessions were conducted in German and participant quotes were translated.

\subsubsection*{Procedure}

The tests were performed on an iPhone 15 Pro, preloaded with two versions of \textit{PPEmployee}, each containing one of the feedback systems.
Throughout the sessions, one author sat beside the participants with a laptop, ready to answer clarification questions and take observational notes.
The user testing started by welcoming the participants, explaining the purpose of the study, informing them of their right to withdraw at any time, and obtaining their consent to participate, following the guidelines of our institution’s Ethics Committee.
Each participant was then asked to complete the following four feedback tasks:
\begin{description}
    \item[Scenario 1 (Bug Report)] \textit{``Make a phone call to the concierge and report any issues that you encounter.''}  
    
    (In this scenario, the app crashes when attempting to initiate the phone call.)

    \item[Scenario 2 (Bug Report)] \textit{``Add your employee pass to Apple Wallet and report any encountered issues.''}  
    
    (In this case, a loading spinner appears briefly without adding the pass to the wallet.)

    \item[Scenario 3 (Feature Request)] \textit{``You are creating a to-do and missing the option to add attachments. Report this feature request to the developer.''}

    \item[Scenario 4 (Feature Request)] \textit{``You wish to have a list of useful phone numbers within the app. Report this feature request to the developer.''}
\end{description}

\begin{table}
\centering
\def\arraystretch{1.4}
\begin{tabular}{|c|p{7.5cm}|}
\hline
\multicolumn{2}{|l|}{\textbf{After completing all scenarios (per system)}} \\
\hline
Q1 & How easy did you find the process of submitting your feedback? \\ 
\cline{2-2}
   & \textbf{Scale:} Very difficult (1) – Somewhat difficult (2) – Somewhat easy (3) – Very easy (4) \\
\hline
Q2 & How helpful do you find the system for submitting your feedback? \\
\cline{2-2}
   & \textbf{Scale:} Not helpful (1) – Somewhat helpful (2) – Helpful (3) – Very helpful (4) \\
\hline
\multicolumn{2}{|l|}{\textbf{At the end of the session} (Responses given verbally)} \\
\hline
Q3 & Explain which feedback system you prefer. \\
\hline
Q4 & What would you like to change about the new system? \\
\hline
Q5 & What concerns do you have and what other comments would you like to share about the new system? \\
\hline
\end{tabular}
\caption{Survey questions to evaluate participants’ perceptions of ease, helpfulness, and overall experience when comparing the traditional feedback system and \approach.}
\label{tab:user-survey}
\end{table}

To encounter the defects during the testing, we reintroduced code from previous versions, as the issues had already been fixed in the current release.  
Participants completed four scenarios with both systems.
After finishing each system, they rated their experience by answering two questions on a four-point semantic scale (Q1 and Q2 in Table~\ref{tab:user-survey}).
At the end, they completed a short exit survey with open questions about their impressions and suggestions for improvement (Q3–Q5).

\subsection{Expert Assessment (RQ2)}

After the user testing, we collected 54 feedback reports, eight from each participant, as they completed four tasks using both methods.
One report was lost due to a technical issue during submission via \approach; therefore, we removed it together with its counterpart.
In total, participants created 28 bug reports from Scenarios 1 and 2, as well as 26 feature requests from Scenarios 3 and 4.

Two experts conducted the evaluation independently: an iOS developer with seven years of professional experience (author, not the app’s developer) and an independent UX specialist with over six years of experience.
This pairing reflects two common perspectives in development teams, a technical one and a design one.
Bug reports were evaluated following the approach of Song et al.~\cite{song2022toward}, who also assessed bug reports collected through their interactive in-app feedback framework. 
Their evaluation was based on the bug quality model proposed by Chaparro et al.~\cite{chaparro2019assessing}.
Each report was evaluated on the availability of three dimensions: \textit{Observed Behavior}, \textit{Expected Behavior}, and \textit{Steps to Reproduce}.
A coding guide~\cite{replicationPackage}, adapted to the two bug-report scenarios, was collaboratively developed by the experts and applied consistently across all reports.
Feature request quality was assessed using the framework of Heck and Zaidman~\cite{heck2017framework}, which provides structured criteria for evaluating the quality of ``just-in-time'' requirements such as feature requests.
We applied all criteria that the framework designates as applicable at creation time, except those not applicable to this user testing, such as ``navigable links''.
The two experts rated each criterion according to the detailed descriptions in the original framework, so no additional coding guide was required.

%% file: chapters/05_evaluationResults.tex
\section{Evaluation Results} \label{sec:eval-results}

\subsection{User Testing (RQ1)}

\begin{table}
    \centering
    \caption{User ratings of the traditional feedback system (text field) versus \approach.}
    \begin{tabular}{|c|cc|cc|c|}
        \hline
        \multirow{2}{*}{Participant} & \multicolumn{2}{c|}{Traditional} & \multicolumn{2}{c|}{\approach} & \multirow{2}{*}{Preference} \\
        \cline{2-5}
         & Ease & Helpfulness & E. & H. &  \\
        \hline
        P1 & 3 & 1 & 3 & 4 & New \\
        P2 & 3 & 2 & 3 & 3 & New \\
        P3 & 4 & 1 & 4 & 3 & New \\
        P4 & 3 & 1 & 4 & 4 & New \\
        P5 & 3 & 1 & 4 & 4 & New \\
        P6 & 2 & 1 & 4 & 3 & New \\
        P7 & 2 & 1 & 4 & 3 & New \\
        \hline
        \textbf{Avg.} & \textbf{2.86} & \textbf{1.14} & \textbf{3.71} & \textbf{3.43} & \\
        \hline
    \end{tabular}
    \label{tab:eval_user_feedback}
\end{table}

The results of the user testing are shown in Table~\ref{tab:eval_user_feedback}.  
For the traditional text based system, participants rated the average ease of reporting feedback at $2.86/4$, while they rated the perceived helpfulness of the system at only $1.14/4$, near the bottom of the scale.  
Participants explained their low ratings with statements such as ``I have to think a lot to formulate an issue report'' and ``The effort of writing so much into this text field makes me angry compared to how simple it was with the other system''.  
For \approach, the average ease of reporting feedback was rated $3.71/4$, while the perceived helpfulness reached $3.43/4$, more than three times higher than the traditional system.  
Participants attributed their positive experience to the system’s intuitive guidance and low effort, stating, for example, ``It's really helpful that reporting starts where I am, so I don't need to explain everything'' and ``Wow, the system seems to understand exactly what I mean''.  
On average, the ease score on reporting feedback increased by $+0.85$, and the helpfulness score tripled ($+2.29$), indicating a clear improvement in user experience with \approach compared to the traditional feedback method.

Nevertheless, the user tests revealed possible improvements for our iOS implementation of \approach.  
First, all participants mentioned that speed gains would help the system become easier to use, as the loading times ranged from 10 to 20 seconds per request.  
Also, as \approach was implemented with a shake-to-report gesture, several participants suggested making this feature more noticeable, which suggests that a simple ``Report Feedback'' button could be more intuitive.  
Further comments suggested that they would have liked an indication of whether the feedback would be sent anonymously or not.  
When sharing positive impressions, participants emphasized that they liked how \approach dynamically adapted to their input and that they did not need to explain their entire situation, as the system already understood the context.  
Observations of user interactions showed that the initial feedback predictions were not always precise.
However, when the model’s suggestions matched the user’s intent ($\sim30\%$ of the time), participants reacted with clear enthusiasm and satisfaction.  
    
In conclusion, to answer \textit{RQ1}, all participants preferred \approach over the traditional text field-based system.  
They perceived \approach as easier to use and much more helpful, while recommending improvements such as faster loading times.

\subsection{Expert Assessment (RQ2)}

\subsubsection*{Bug Reports}

\begin{table}
\centering
\caption{Average bug report quality ratings (n=28; Traditional System n=14; \approach n=14). Ratings were given on a 3-point scale (0,1,2).}
\label{tab:bug_quality}
\begin{tabular}{|p{4.2cm}|r|r|}
\hline
Quality Dimension & Traditional & \textbf{\approach} \\
\hline
Observed Behavior   & 0.93 & 1.50 \\
Expected Behavior   & 2.00 & 2.00 \\
Steps to Reproduce  & 0.14 & 2.00 \\
\hline
\end{tabular}
\end{table}

We present the results of the bug report quality assessment in Table~\ref{tab:bug_quality}.
Overall, the reports generated through \approach were rated higher in quality than those written with the manual feedback system.
Generally, reports created with \approach included more detailed descriptions along with contextual data, which resulted in the higher score of $1.50/2$ for \textit{Observed Behavior}.
In contrast, the manually written reports in the traditional system were often brief and lacked essential details.
Participants entered short messages such as ``App crash when Concierge was called'', without specifying precisely when the crash occurred.
Consequently, the experts rated these reports as adequate rather than perfect in the \textit{Observed Behavior} dimension, as they did not fully explain the situation, for example, that the crash happened immediately after pressing the call concierge button.
\approach automatically captured \textit{Steps to Reproduce} through interaction logging, which explains its perfect rating ($2/2$).
In the traditional system, participants never mentioned reproduction steps explicitly, resulting in a near-zero rating.
For \textit{Expected Behavior}, both systems achieved the maximum score, as experts could easily infer what should have happened in these simple scenarios, e.g., the app should not crash.
In summary, \approach produced substantially more complete and structured bug reports by combining the summaries with contextual data such as device information, interaction logs, and follow-up question responses.
This led to higher expert ratings for \textit{Observed Behavior} and \textit{Steps to Reproduce}, where the traditional reports frequently lacked necessary details.

\subsubsection*{Feature Requests}

\begin{table}
\centering
\caption{Average feature request quality (n=26; Traditional n=13; \approach n=13).}
\label{tab:fr_quality}
\begin{tabular}{|l|r|r|}
\hline
Quality Dimension & Traditional (\%) & \textbf{\approach (\%)} \\
\hline
Summary                   & 100.0 & 100.0 \\
Description               &   0.0 & 100.0 \\
Specify problem            &   0.0 &  53.8 \\
Correct summary            &  69.2 &  92.3 \\
Product version            &   0.0 & 100.0 \\
Correct language           &  92.3 & 100.0 \\
Atomic                     & 100.0 & 100.0 \\
Glossary                   & 100.0 & 100.0 \\
All comments necessary     &  99.2 &  99.6 \\
\hline
\end{tabular}
\end{table}

We present the results of the feature request quality assessment in Table~\ref{tab:fr_quality}.  
\approach achieved perfect ratings for summary and description, as each report includes a summary and the remaining content was categorized as description.  
In contrast, users in the traditional text-field system submitted only brief statements, which the experts evaluated solely as summaries, resulting in $0\%$ for description.  
The dimension \textit{Specify problem} received a lower rating for both approaches, with $0\%$ for the traditional method and $53.8\%$ for \approach.  
One user report was ``Add attachments to To-Dos'', which only stated a desired solution without explaining the underlying problem.  
As a comparison, a feature request generated through \approach contained the user intent summary: ``I want to be able to attach photos and various files to to-dos to make sharing important information effortless'', which not only specifies the desired feature but also describes the underlying problem.  
However, the experts noted that the follow-up questions in \approach often focused more on refining a solution rather than uncovering the underlying issue.
During the assessment, the UX expert pointed out that this difference reflects a common challenge observed in the real-world: users often describe a solution rather than their problem.
The expert emphasized that this can be counterproductive, as designing the appropriate solution is part of the expert’s role and relies on their professional experience.
Users, on the other hand, may not always know what they truly need \cite{maalej2015bug}.
The dimension \textit{Correct summary} was higher for \approach ($92.3\%$) than for the traditional method ($69.2\%$), as the AI-assisted reports were more specific, whereas traditional ones were often vague, such as ``Phone book with emergency numbers''.
Finally, the product version was not included in the traditional reports, resulting in $0\%$, while \approach automatically captured this information, achieving $100\%$.  
Both approaches scored well on the remaining criteria, such as \textit{Correct language} and \textit{Atomic}.

\subsubsection*{Inter-Rater Agreement}
Across both assessments, the two experts showed a very high level of consistency in their independent evaluations.  
Cohen’s Kappa, a widely used measure of inter-rater reliability \cite{sun2011meta}, was $\kappa = 0.906$ for the bug reports and $\kappa = 0.9207$ for the feature requests, which according to the interpretation benchmarks by Altman \cite{altman1990practical} and Landis and Koch \cite{landis1977measurement} indicates ``almost perfect'' agreement.  
Where differences occurred, they were mainly due to missed or overlooked details.  
After the individual assessments, the experts discussed these few cases and reached a mutual consensus on the final ratings.
\\ \\
To answer \textit{RQ2}, expert evaluations showed that bug reports and feature requests generated with \approach were more complete than those created with the manual feedback system.
However, the experts also noted that some follow-up questions tended to focus more on refining solutions rather than uncovering the underlying problems, highlighting potential to adjust \approach to focus on identifying root causes of user issues.

%% file: chapters/06_threats.tex
\section{Threats to Validity} \label{sec:threats}

As for all empirical studies, our work is subject to several potential threats to validity, which we briefly discuss below.

\subsection{Internal and External Validity}

As is common in user studies such as usability testing, several forms of bias can occur, including observer bias or social desirability bias, where participants provide responses they believe please the observer, for example, due to a positive relationship \cite{natesan2016cognitive}.  
Since we conducted a within-subject user testing, participants may have reacted differently when reporting the same issue twice, for example by giving shorter or less detailed reports, though counterbalancing was applied to mitigate this effect.
During expert assessment, ratings were performed by two evaluators, one of whom was an author of this paper, which may introduce subjective bias.  
To mitigate this, we used detailed annotation guidelines and independent peer evaluation.  

Similar to case studies, our results are not intended for broad generalization but rather to explore insights into the feasibility of the \approach approach \cite{stol2018abc}.  
Further studies are needed to explore how MLLM-guided user feedback processes might behave in other contexts, as our study focused on a single app from a local gym with only a limited number of participants.
Examining broader behavioral patterns of users would require a different sample size.  
We consider the sample size of our second study appropriate for its purpose, as it focuses on a comparative analysis of 54 feedback reports.

The traditional approach used for comparison with \approach consisted of a simple text field.
Other apps may implement more complex or structured feedback mechanisms, differing in both design and usefulness.
Additionally, our testing scenarios included only two bug reports and two feature requests, which do not capture the full diversity of possible feedback types such as content or usability suggestions \cite{pagano_user_2013, nielsen1994usability}.  
To improve generalizability, future studies should replicate our work across multiple app categories, feedback types, and user populations.  

\subsection{Construct Validity}

During expert evaluation, we applied two distinct metrics from prior research, tailored to bug reports and feature requests respectively.
While these metrics capture important aspects of feedback quality, alternative measures could yield different insights.  
For bug reports, we employed the quality model by Chaparro et al.~\cite{chaparro2019assessing}, following prior work by Song et al.~\cite{song2022toward}, which evaluates three elements: \textit{Observed Behavior}, \textit{Expected Behavior}, and \textit{Steps to Reproduce}.  
Alternative metrics, such as those proposed by Bettenburg et al.~\cite{bettenburg2008makes}, emphasize different attributes (e.g., stack traces or test cases), which could lead to varying results; although the discussed advantages of \approach would likely still hold.

For feature requests, we applied the quality framework by Heck and Zaidman~\cite{heck2017framework}, focusing on the ``at creation'' criteria that evaluate quality at the time a request is submitted.  
Their framework distinguishes between ``at creation'' and ``just-in-time'' criteria, where the latter are expected to hold later in the development process.  
Accordingly, we limited our evaluation to elements applicable at submission time, while noting that certain ``just-in-time'' aspects such as including a rationale or identifying the originating screen could also provide value in future assessments.  
Furthermore, alternative metrics, such as the framework by Wouters et al.~\cite{wouters2022crowd} originally designed for user issues, could have also been adapted for evaluating feature requests, which could have led to varying results.

Another potential construct threat is that our feedback scenarios were predefined.  
Although they were derived from real feedback examples, participants may still not have expressed themselves as naturally as they would in a real usage setting.  

Finally, our study used OpenAI's GPT-4.1 as the underlying MLLM.  
Different models such as Gemini or Claude, as well as different prompting strategies, might produce varying outcomes.
This limitation, however, does not undermine the insights derived from our analysis under the chosen settings, and the conceptual foundation of \approach.
Moreover, recent reports indicate that the performance of leading models is converging due to similarities in model architectures and overlap in training data \cite{maslej:AIIndex:2025}.

%% file: chapters/07_relatedWork.tex
\section{Related Work} \label{section:related_work}

Prior work on app feedback systems has focused on capturing context-rich reports within apps, whereas complementary research investigates how conversational or automated follow-up approaches can elicit missing details from users.
We summarize both lines of work and discuss \approach as an integrated, context-aware framework that combines in-app feedback capture with adaptive conversational elicitation.

\subsection{App Feedback Systems}

Prior studies show that App Store reviews are a main channel for developers to collect user feedback \cite{Hassan:EMSE:2018, pagano_user_2013}, in addition to other channels such as social media \cite{Martens:RE:19} and product forums \cite{Tizard:RE:2019}.
Such reviews help identify issues and guide app improvements, supporting user-driven quality evaluation \cite{maalej2025automated}.
However, the quality of this feedback varies widely \cite{maalej2025automated}, with up to 75\% of reviews in the App Stores consisting mainly of praise rather than actionable content \cite{pagano_user_2013, Martens:RE:19}.
Different research approaches have been proposed to help users provide valuable feedback for developers.
Seyff et al.~\cite{seyff2014appecho} presented AppEcho in 2014, an early in-app feedback approach that enables users to capture feedback in situ, directly when an issue occurs, including contextual information. 
The system allows users to take screenshots and submit short text or audio feedback within the to be reported app itself. 
Their early study highlighted the value of contextual, in-app feedback compared to traditional channels outside the app.
In their research on contextual information contained in user feedback on social media, Martens and Maalej \cite{Martens:RE:19} sketched the idea of chatbots interacting with users to gather additional contextual information.
The closest approach to \approach was introduced by Song et al.~\cite{song2022toward}, who proposed BURT, a web-based, rule driven chatbot for interactive bug reporting on Android.
BURT guides users through the reporting process using natural language to ensure the three key elements of the bug quality model by Chaparro et al.~\cite{chaparro2019assessing} (\textit{Observed Behavior}, \textit{Expected Behavior}, and \textit{Steps to Reproduce}) are included.
It validates report quality by matching user inputs against a precomputed app execution model built from automated GUI exploration and crowdsourced traces.
While BURT relies on rule driven parsing and part of speech tagging for natural language understanding,
\approach leverages an MLLM to interpret user input and generate context-aware follow-up questions based on live app context.
This enables \approach to support a broader range of feedback types beyond bugs and to create adaptive interactions that better capture relevant feedback information, capabilities that may be difficult to achieve with a rule based system.
Hence, \approach extends their foundational idea through the integration of GenAI, which enables new capabilities through stronger contextual understanding in app based feedback scenarios \cite{Pourasad:DoesGenAIMake:2025}.

\subsection{User-Developer Conversations}

When users report feedback, they often omit crucial details needed by developers to understand, reproduce, or fix issues.  
Prior studies show that developers frequently engage in clarification dialogues to gather such missing information.  
Hassan et al.~\cite{Hassan:EMSE:2018} found that in about 45\% of developer responses to App Store reviews, developers asked users for more details, often redirecting them to other channels such as email.  
Martens and Maalej~\cite{Martens:RE:19} observed that non-technical users often provide unstructured feedback and omit essential contextual details such as steps to reproduce or environment information.  
Breu et al.~\cite{breu2010information} further reported that many users never reply to follow-up questions, and those who do often take several hours or days.  
Together, these studies highlight that user feedback is valuable but often incomplete, leading to time-consuming clarification exchanges.

Researchers have explored interactive methods to elicit missing information from users for improving feedback completeness.
Imran et al.~\cite{imran2021automatically} presented Bug-AutoQ, which supports developers when facing incomplete bug reports from users by automatically recommending follow-up questions that developers can choose and send to users for clarification.  
It retrieves and ranks suitable questions from similar historical reports using information retrieval and neural ranking models.  
This approach, however, still depends on users to provide the missing information after initial feedback submission, whereas \approach collects it interactively during the feedback process.  
BugMentor~\cite{mukherjee5000516bugmentor} complements this work by automatically generating answers to developers’ questions on bug reports.
It produces concise answers using structured information from existing bug reports through structured information retrieval and neural text generation.  
This approach, however, requires users to have already provided detailed bug reports, while \approach ensures that the necessary information is gathered the moment users report feedback.

More recently, Shen et al.~\cite{shen2025requirements} demonstrated that LLMs can generate effective follow-up questions in real time during requirements elicitation interviews, suggesting the potential of interactive GenAI systems to support information elicitation from stakeholders, similar to \approach.
Kuric et al.~\cite{kuric2025unmoderated} investigated using LLMs in usability testing, where GPT-4 generated adaptive follow-up questions during unmoderated sessions to elicit more detailed participant responses.  
This work demonstrates the potential of LLM-powered conversational systems in another domain to collect richer qualitative data, aligning closely with the goals of \approach.

%% file: chapters/08_discussionAndConclusion.tex
\section{Summary and Future Work} \label{sec:conclusion}

In this paper, we proposed a novel approach to address the persistent gap between what app users report and what developers need \cite{maalej2025automated,Martens:RE:19,Hassan:EMSE:2018,breu2010information}.
\approach guides users through an adaptive in-app feedback reporting process to enable easier context-aware feedback reporting while submitting more complete reports to developers.
\approach leverages MLLMs to dynamically adjust follow-up questions based on live app context, making it applicable across different feedback types and app domains.

We conducted a user testing study where users of a gym app reported feedback using \approach and the app's conventional feedback system. 
Participants rated \approach as easier and more helpful to use than the conventional feedback form.
The 54 resulting feedback reports were evaluated by two industry experts, also revealing that \approach generated higher quality bug reports and feature requests.
These findings indicate that MLLM-guided, context-aware feedback systems can improve both user experience and developers’ insights by simplifying the feedback reporting and increasing the completeness of feedback reports, thus reducing the need for asynchronous follow-up clarifications.

We see several directions for future work.
First and foremost, we think that \textbf{privacy protection} is a major future direction, as utilizing vendor LLMs always carries a significant potential risk due to privacy and security concerns \cite{wu2024unveiling}.  
Sending sensitive user data, even with consent, poses challenges, for example when screenshots or logs unintentionally contain personally identifiable or financial information \cite{feretzakis2024privacy}.  
The current implementation of \approach requires users’ informed consent and should be advanced by adding techniques for masking or obfuscating sensitive data to preserve user privacy \cite{kotey2024preserving, MiraouiWoringer2024}.  
However, we think that a more promising line emerges from recent advances in efficient on-device models.
Those suggest that running models locally on users’ devices is increasingly feasible, as vendors such as Apple are heavily investing in optimized on-device hardware to enable real-time, private model inference \cite{vasu2025fastvlm, apple_coreml_framework}.
Powerful on-device processing that generates high-quality feedback reports, while keeping context data and user interactions local, could provide strong functionality with minimized risk.
Nevertheless, a proper filtering layer would still be needed before sending the final report to the developer to ensure privacy compliance \cite{kotey2024preserving}.  
Future research should focus on building privacy-preserving pipelines for interactive feedback collection, as this remains a key challenge across many user-facing GenAI systems \cite{wu2024unveiling, feretzakis2024privacy, duffourc2023privacy}.

The generalizability of our results is limited, since they aim to provide initial insights into a GenAI-powered feedback reporting approach.
Therefore, we encourage future research to replicate and extend our study by augmenting the dataset with additional apps and a wider range of feedback cases, and by further refining the \approach concept and its underlying implementation. 
Future evaluations could also examine how different models or prompt configurations perform for distinct feedback types, such as bug reports, feature requests, or usability issues. 
Additional studies could focus on understanding the role and effectiveness of different follow-up question strategies in improving feedback quality; for instance, by building on frameworks and taxonomies for characterizing follow-up questions \cite{kundu2020learning, ge2022should, zimmermann2025questions, mansilla2024taxonomy}.

To conclude, our findings align with the direction suggested by Song et al.~\cite{song2022toward}, implying that feedback collection may be moving from static input forms toward increasingly interactive experiences. 
Our work extends this idea through GenAI-driven feedback reporting, enabling context-aware follow-up clarifications while reducing the feedback overhead and maximizing its value for both users and developers.